\documentclass[12pt]{article}
\usepackage{amssymb,amsmath,eucal}
\begin{document}

\begin{titlepage}

                            \begin{center}
                            \vspace*{5mm}
\Large\bf{Nilpotence and the generalized uncertainty principle(s)}\\

                            \vspace{2.5cm}

              \normalsize\sf    NIKOS \  \  KALOGEROPOULOS $^\dagger$\\

                            \vspace{2mm}
                            
 \normalsize\sf Weill Cornell Medical College in Qatar\\
 Education City,  P.O.  Box 24144\\
 Doha, Qatar\\

                            \end{center}

                            \vspace{2.5cm}

                     \centerline{\normalsize\bf Abstract}
                     
                           \vspace{3mm}
                     
\normalsize\rm\setlength{\baselineskip}{18pt} 

\noindent  We point out that some of the proposed generalized/modified uncertainty principles
originate from solvable, or nilpotent at appropriate limits, ``deformations" of Lie algebras. 
We briefly comment on formal aspects related to the well-posedness of one of these algebras. 
We point out a potential relation of such algebras with Classical Mechanics in the spirit of the
symplectic non-squeezing theorem. We also point out their relation to a  hierarchy of generalized 
measure theories emerging in a covariant formalism of quantum gravity.     \\

                             \vfill

                             
                             \vfill

\noindent\rule{8cm}{0.2mm}\\
   \noindent \small\rm $^\dagger$  E-mail: \ \  \small\rm nik2011@qatar-med.cornell.edu\\

\end{titlepage}


                                                                                 \newpage

 \normalsize\rm\setlength{\baselineskip}{18pt}

                     \centerline{\large\sc 1. \  \  Introduction}

                                                                                \vspace{5mm}

The possibility that the Heisenberg uncertainty principle is modified by quantum gravitational effects has been
first proposed almost a half century ago [1]. More recently, this idea resurfaced [2]-[8], mostly motivated by 
a wish to naturally incorporate a minimal length [9], [10] in the various approaches to quantum gravity. 
Such a generalisation is warranted close to the Planck scale, around which the Compton wavelength of a particle
becomes comparable to its Schwarzschild radius. There has been
a veritable explosion of interest in this topic during the last two decades, during which formal variations [11], 
and their statistical mechanical and phenomenological implications [12]-[18] have been being explored. 
Implications of the generalised uncertainty principles for quantum field and gauge theories have also recently emerged [19]-[21].\\

It is probably not too surprising that there is no ``unique", ``natural" or even ``best" generalisation of the Heisenberg 
uncertainty principle. Such a generalisation really depends on the goals that one wishes to attain, and is ultimately 
justified by its predictions or by a physical principle, already known or new, that may be uncovered lying at its foundations.
As a result, one encounters several versions of the generalised uncertainty principle, which stem from different
generalisations of the Heisenberg algebra. Ultimately, such generalised uncertainty principle should arise and be justified by 
a theory of quantum gravity. Since such a universally acceptable theory is currently lacking several paths, mostly 
phenomenologically motivated,  have been taken toward the formulation of generalised  uncertainty principles [2]-[18].\\

In the present work, we examine some formal aspects of such proposed generalisations of the Heisenberg algebra.  
In Section 2, we notice that most such proposed algebras are solvable, and even nilpotent in appropriate limits, deformations of Lie algebras.
We also comment on why they may fail to have these properties.  Such structures interpolate between the Heisenberg algebra, 
which is 2-step nilpotent, hence solvable and the full structure of a  potential non-commutative geometry [22],[23] 
that may be used to quantize gravity. Some technical points pertaining to some of these algebras are also very briefly addressed. 
In Section 3, we make some remarks pertaining to the generalised algebras
as seen through the ``classical" symplectic non-squeezing theorem [25]-[35] and their ``counterpart" in the covariant formalism 
which is  a generalized measure theory [36]-[38]. 
Section 4, presents an outlook and speculations. In the Appendix we state some very well-known facts from real and Fourier analysis [39]-[41] 
as well as the theory of pseudo-differential operators [42], aiming at making our exposition somewhat more self-contained. \\

                                                                                             \vspace{10mm}
  
 
                                                     \centerline{\large\sc 2. \  Nilpotence and the Generalized  Uncertainty Principle(s)}
                                                     
                                                                                           \vspace{5mm}
                                                                                                           
\noindent{\large\bf 2.1} \  \ The concepts of nilpotent and solvable groups and algebras  are central in the structure and classification of both discrete and 
``continuous"  groups and algebras [43]. Let \ $G$ \ indicate a group with elements \ $g_I$ \ where \ $I$ \ is a discrete or continuous index set. 
The (group) commutator is the subgroup indicated by \ $[G, G]$ \  having elements
\begin{equation}
     [G, G] = \left\{ g_i^{-1}g_j^{-1} g_i g_j, \ \ \ \forall \ i,j \in I \right\}
\end{equation}
In a similar manner, when one considers two subgroups \ $H_1, H_2 \leq G$, \ with elements \ 
$H_1 = \{ h_{1j}, \ j\in J \}, \ \ H_2 = \{ h_{2k}, \  k\in K \}$ \ with \ $J, K$ \ subsets of $I$, then their commutator subgroup \ $[H_1, H_2]$ \ 
is given by
\begin{equation}
      [H_1, H_2] = \left\{ h_{1j}^{-1} h_{2k}^{-1} h_{1j} h_{2k}, \ \ \ \forall \ j\in J, \ k\in K \right\}
\end{equation}

Consider the following commutator groups defined iteratively by 
\begin{equation}
G_{(1)} = G, \ \ \ \ G_{(i+1)} = [G, G_{(i)}], \ \ \ \  i\in \mathbb{N} 
\end{equation}
The descending central series is
\begin{equation}
  G \geq G_{(2)} \geq G_{(3)} \geq \ldots
\end{equation}
A group \ $G$ \ is called $n$-step nilpotent, if its lower central series terminates after $n$-steps, 
namely if there is \ $n\in \mathbb{N}$ \ such that  \ $G_{(n+1)} = 1$. \ Several other equivalent definitions exist for nilpotent groups. 
Examples of nilpotent groups: all Abelian groups are 1-step nilpotent. The Heisenberg group is 2-step nilpotent. By contrast, the quaternion and the 
rotation groups are not nilpotent.\\ 

Consider the following commutator groups  defined iteratively by
\begin{equation}
   G^{(1)} = G, \ \ \ \ G^{(i+1)} = [G^{(i)}, G^{(i)}], \ \ \ \ i\in\mathbb{N} 
\end{equation}
The derived series is
\begin{equation}
    G \geq G^{(2)} \geq G^{(3)} \geq \ldots
\end{equation}
A group \ $G$ \ is solvable if its derived series terminates after $n'$ steps, namely if there is \ $n'\in\mathbb{N}$ \ such that \ $G^{(n'+1)} = 1$. \  
Examples of solvable groups:  all Abelian groups are solvable. More generally, all nilpotent groups are solvable, as can be readily seen. 
A solvable but non-nilpotent group is the symmetric group of 3 elements \ $S_3$. \
 For discrete groups, the Feit-Thompson theorem states that every finite group of odd order is solvable.\\

For a Lie algebra \ $\mathfrak{g}$ \ similar definitions apply, by using the matrix instead of the group commutator. 
Then Engel's theorem states  that a Lie algebra is nilpotent if the adjoint map \  $ad_x (y) = [x, y], \ x, y \in \mathfrak{g}$ \ is a nilpotent operator, 
namely if there is  $n\in\mathbb{N}$  such that \ $ad(x)^n = 0, \ \ \forall \ x\in\mathfrak{g}$. \ Moreover, \ $\mathfrak{g}$ \ is solvable 
if and only if  \ $[\mathfrak{g}, \mathfrak{g}]$ \ is nilpotent. The notation for Lie algebras \ $\mathfrak{g}$ \ that we will use is analogous to 
that for groups, as given above. Analogous definitions can be used for associative algebras endowed with commutators as will be done in the sequel.\\


\noindent{\large\bf 2.2} \  \ One can easily see that the Heisenberg algebra of Quantum Mechanics, giving rise to the ``ordinary" uncertainty principle, is 2-step nilpotent, hence solvable, since 
\begin{equation}
     [x_i, p_j] \  = \ i\hbar\delta_{ij}, \ \ \ \ i, j = 1, \ldots, n
\end{equation}
and all other commutators are zero, where the dimension of the phase space \ $\mathcal{M}$ \ is \ $2n$. \ 
Then 
\begin{equation}
    [x_i, [x_j, p_k]] = [p_i, [x_j, p_k]] = 0 
\end{equation}
with the other 2-step commutators trivially zero. Now consider the n-dimensional
rotationally symmetric Kempf-Mangano-Mann (KMM) deformation [7],[8]  of the Heisenberg algebra given by
\begin{equation} 
    [x_i, p_j] \ = \ i\hbar  (1 + \beta^2 \|p\|^2) \delta_{ij} 
\end{equation}
\begin{equation}    
    [p_i, p_j] \ = \ 0 
\end{equation}
\begin{equation}    
    [x_i, x_j] \ = \ 2i\hbar \beta^2 (p_i x_j - p_j x_i) \\
\end{equation}
where \ $\beta\in \mathbb{R}_+$ \ and 
\begin{equation}
\|p\|^2 = \sum_{i=1}^n \ p_i p_i
\end{equation}
We immediately observe that all elements of the second step of the derived series \ $\mathfrak{g}^{(2)}$ \ 
are zero, except  the following ones that require some straightforward calculations
\begin{equation}
   \left[ [x_i, p_j], [x_k, x_l]  \right] = 0 
\end{equation}
and 
\begin{equation}
   \left[ [x_i, x_j], [x_k, x_l]\right] = 2 (i\hbar \beta)^2 (i\hbar) \left(1 + \beta^2 \|p\|^2\right) 
                   \left(\delta_{il} [x_j, x_k] + \delta_{jk} [x_i, x_l] + \delta_{jl} [x_k, x_i] + \delta_{ik} [x_l, x_j]\right)
\end{equation}
Using those, one can express the elements of the higher elements in the derived series in terms of those of \ $\mathfrak{g}$ \ and \
$\mathfrak{g}^{(2)}$. \ As can be immediately seen, the derived series does not terminate, in general. \\

We proceed by further simplifying matters, in order to get a firmer control of the algebra. An obvious way to achieve this goal  is to impose conditions 
that make \ $\mathfrak{g}^{(2)}$ \ Abelian. One way to attain this  is to consider only the ``semi-classical limits" \ $\hbar \rightarrow 0$ \ or \ $\beta 
\rightarrow 0$, \ or both, of the KMM deformation. Such a ``Inon\"{u}-Wigner"-like contraction is implemented by ignoring all terms that are of quadratic 
or higher order in \ $\hbar$  \  and of quartic or higher order in \ $\beta$. \ A second way to proceed is by foregoing altogether all traces of non-commutativity between the ``spatial" variables \ $x_i$ \ 
by imposing 
\begin{equation}
       [x_i, x_j] = 0
\end{equation}
Obviously, (15) is a significant simplification of the KMM deformation. It is adopted by the ``modified uncertainty principle" as will be seen in the sequel. 
If either of these simplifications are made, then the corresponding subalgebra of the KMM deformation is 2-step solvable as can be seen from (14).
We have to be somewhat careful though. If we assume (14), and we omit terms of quadratic and higher order in
\ $\hbar$ \ then what remains is the Heisenberg algebra, so we get nothing new. Hence, to get a nontrivial result, we are forced, in addition to (14)
to omit only terms of quartic or higher order in \ $\beta$. \ By using this approximation, we go beyond the Heisenberg algebra, since
\begin{equation}
            \begin{array}{ll}
                           [x_l, [x_k, [x_i, p_j]]] \ = &  2(i\hbar)^3 \beta^2 (1 + \beta^2 \|p\|^2) \delta_{ij} \delta_{kl} \  + \\ 
                                                                  &  \ \  + \ 2 (i\hbar)^3 \beta^4  \delta_{ij}  \{ (1 + \beta^2 \|p\|^2) \|p\|^2 \delta_{kl} + 2 (1+ \beta^2 \|p\|^2) p_k p_l \}
            \end{array}
\end{equation} 
reduces to 
\begin{equation}
    [x_l, [x_k, [x_i, p_j]]] = 2(i\hbar)^3 \beta^2 \delta_{ij} \delta_{kl}
\end{equation}
which is obviously an element of  the centre of \  $\mathfrak{g}$. 
Then all  4-step commutators, i.e. all elements of \ $\mathfrak{g}_{(5)}$ \ are trivial. In other words, under the above approximations, 
the KMM deformation (9)-(11) reduces to a 4-step nilpotent algebra.  \\


\noindent {\large\bf 2.3} \  \ Maggiore [3]-[5] proposes a generalization of the Heisenberg uncertainty relations that can be derived from the Lie algebra
having generators \ $x_i, \ p_j, \ J_k, \ \  i,j,k = 1,2,3$ \ which obey the commutation relations
\begin{equation}
    [x_i, x_j] \ = \  - i \ \frac{\hbar^2}{\kappa^2} \ \epsilon_{ijk} \ J_k
\end{equation}
\begin{equation} 
   [x_i, p_j] \ = \  i\hbar \ \sqrt{1 + \frac{E^2}{\kappa^2} } \ \delta_{ij}
\end{equation}
\begin{equation}
   [p_i, p_j] \ = \ 0 
\end{equation}
\begin{equation}
   [J_i, x_j] \ = \ i\epsilon_{ijk} x_k
\end{equation}
\begin{equation}
   [J_i, p_j] \ = \  i\epsilon_{ijk} p_k
\end{equation}
\begin{equation}
  [J_i, J_j] \ = \ c_{ijk} \ J_k
\end{equation}
where \ $J_i, \ i=1,2,3$ \ stand for the components of the total angular momentum operator, \ $c_{ijk}$ \ are the structure constants of the Lie algebra 
of \ $SU(2)$ \ and \ $\kappa$ \ is the ``deformation" parameter  which is identified with the Planck mass. The essential difference between this 
algebra and  the Heisenberg algebra can be essentially traced back to (18). As \ $\kappa \rightarrow \infty$, \ we recover the direct product  of the 
Heisenberg algebra with the Lie algebra of \ $SU(2)$. \ The latter however cannot become solvable, in any approximation in terms of \ $\kappa$. \
To justify this, consider the Killing form of a Lie algebra \ $\mathfrak{g}$, \ defined as the symmetric bilinear form on \ $\mathfrak{g}$ \ given by
\begin{equation}
      K(x,y) \ = \  tr (ad(x) \ ad(y)), \hspace{8mm} x, y \ \in \mathfrak{g}
\end{equation}
Cartan's criterion states that a Lie algebra \ $\mathfrak{g}$ \ is solvable if and only if its Killing form satisfies
\begin{equation}
     K(x, [y,z]) = 0,  \hspace{8mm} x, y, z \ \in \mathfrak{g}
\end{equation}
 It is straightforward to check that every subalgebra of a solvable Lie algebra is also solvable. 
 Hence if Maggiore's extension could become solvable in some non-trivial (namely, not resulting in the Heisenberg algebra) approximation in terms of
$\kappa$ \ then its \ $SU(2)$ \ subalgebra should also have a degenerate Killing form. This is impossible however as the \ $SU(2)$ \  commutation 
relations do not depend on the value of \ $\kappa$ \ in Maggiore's deformation, as is obvious in (23). Hence Maggiore's deformation cannot give rise to a 
nilpotent algebra either, in some approprite limit in terms of \ $\kappa$. \ We conclude that our approach and subsequent conclusions do not apply to 
Maggiore's generalization of the Heisenberg algebra (18)-(23). \\


\noindent{\large\bf 2.4} \  \ The Das-Vagenas (DV) generalised uncertainty relation is a result of the associative algebra endowed with a bracket given by
\begin{equation}  
  [x_i, p_j] \ = \  i\hbar (\delta_{ij} + \zeta \delta_{ij} \|p\|^2 + 2\zeta p_i p_j),  \hspace{8mm} i, j = 1, \ldots, n 
\end{equation}
\begin{equation}
  [x_i, x_j] \ = \ [p_i, p_j] \ = \ 0, \hspace{8mm} i, j = 1, \ldots, n 
\end{equation}
with \ $\zeta = \zeta_0 / (M_{Pl} c^2)$ \ where \ $M_{Pl}$ \ denotes the Planck mass. This is an anisotropic variation, provided by the term \ 
$2\zeta p_ip_j$ \ of the KMM deformation (9)-(11) with the additional simplification that the spatial coordinates commute as in (15). 
Requiring (15) instead of (11) is a considerable simplification of the KMM proposal, 
conceptually more closely aligned to ordinary rather than to non-commutative geometry. As can be readily seen, 
this algebra is 3-step solvable as all elements of \ $\mathfrak{g}^{(2)}$ \ are zero. On the other hand, for nilpotency we have
\begin{equation}
    [p_k, [p_i, p_j ]] \ = \ 0
\end{equation}
but
\begin{equation}
    [x_k, [x_i, p_j ]] = 2(i\hbar)^2 \zeta (1 + 3 \zeta \|p\|^2) \delta_{ij} p_k + 2 (i\hbar )^2 \zeta 
         \left\{ (1+ \zeta \|p\|^2) (\delta_{ik} p_j + \delta_{jk} p_i) + 2 \zeta p_i p_j p_k )\right\}
\end{equation}
As was also observed in the case of the KMM algebra, the DV one is not nilpotent unless one resorts to some approximations. The most straightforward 
assumption is to consider only terms vanishing as quadratic or higher powers of  \ $\hbar$ \ in which case (29) becomes zero. In this approximation the DV 
algebra is 3-step nilpotent. On the 
other hand, someone may wish to keep only terms  up to second order in $\zeta$. \ Then (29) reduces to
\begin{equation}
   [x_k, [x_i, p_j ]] \ = \ 2(i\hbar)^2 \zeta^2 (\delta_{ij} p_k + \delta_{ik} p_j + \delta_{jk} p_i )
\end{equation}
The only 3-step non-trivial commutator is, in the aforementioned approximation in terms of $\zeta$, 
\begin{equation}
   [x_l, [x_k, [x_i, p_j ]]] \ = \ 2(i\hbar)^3 (\delta_{ij}\delta_{kl} + \delta_{jk}\delta_{il} + \delta_{ik}\delta_{jl})
\end{equation}
which being a central element of \ $\mathfrak{g}$ \ implies that 
\begin{equation}
  [x_m, [x_l, [x_k, [x_i, p_j]]]] = 0
\end{equation}
All the other commutators have been trivially zero from the previous step. We see that the presence of the anisotropic term in (26) does not even affect the 
step at which the DV algebra becomes nilpotent when compared to the KMM case.\\


\noindent{\large\bf 2.5} \  \ The Ali-Das-Vagenas (ADV) ``modified uncertainty principle"  [14], [15] generalizes the spatial-momentum commutator of the 
KMM deformation and extends the DV generalized algebra to 
\begin{equation}
     [x_i, p_j] \ = \   i\hbar \left\{ \delta_{ij} - \alpha \left(  \|p\| \delta_{ij} + \frac{p_i p_j}{\|p\|}  \right) + 
                                              \alpha^2 \left( \|p\|^2 \delta_{ij} + 3p_i p_j \right)  \right\},  \hspace{5mm} i,j = 1, 2, 3  
\end{equation}
The generalisation to \  $n$ \ dimensions is straightforward. 
Here \ $\alpha = \alpha_0 l_{Pl} / \hbar$ \ where \ $l_{Pl}$ \ is the Planck length.
The ADV algebra assumes, as  the DV case (26), (27) above  that
\begin{equation}
     [x_i, x_j] \  = \ [p_i, p_j] \ = \ 0, \hspace{8mm} i, j = 1, 2, 3. 
\end{equation}
In this case the ``deformation" parameter is indicated by \ $\alpha$. \ As in the case of (26), (33) is also 3-step solvable since \ 
$\mathfrak{g}^{(2)}$ \  is also trivial,  as can be seen by a straightforward computation. Notice that due to the fact that the 
canonical momenta commute (34), the two potentially ``dangerous" issues being the exact operator ordering in the fraction of (33), as well as
the exact way that \ $\|p\|$ \ and \ $\frac{1}{\|p\|}$ \ are defined, can be temporarily ignored.  \\

To check the nilpotency of the ADV algebra, we will work at a formal level, leaving potential justifications of these steps for Subsections  2.6, 2.7 
in the sequel.
Consider an operator of interest, let's say \ $p_i$. \ Then define its inverse \ $\frac{1}{p_j}$ \ by demanding 
\begin{equation}
     p_i \ \frac{1}{p_j} \ = \ \delta_{ij} \hspace{10mm} i,j  = 1,\ldots, n
\end{equation}
There is no need to distinguish, naively at least, a left from a right multiplication, because to due to (34), it is expected that both one-sided multiplications
 will  give the same results. We use repeatedly that the commutator is a derivation, as well as (34), and find from (12)
\begin{equation}
    [x_k, \| p\|^2] \ = \ 2 \ \sum_{j=1}^3 \ [x_k, p_l] \ p_l
\end{equation}
With the definition (35) this can be rewritten as 
\begin{equation}
   [x_k, \| p\| ] \ = \ \sum_{l=1}^3 \ [x_k, p_l] \ p_l \ \frac{1}{\| p \|}
\end{equation}
so (35) results in 
\begin{equation}
   [x_k, \frac{1}{\|p\|}] \ = \ - \ [x_k, \|p\|] \ \frac{1}{\|p\|^2} 
\end{equation}
Taking into account (36)-(38), a calculation gives, up to terms of order \ $\alpha^2$, \ that 
\begin{equation}
         \begin{array}{ll}
                 [x_k, [x_i, p_j ]] \ = & -  (i\hbar )^2 \alpha \ \frac{1}{\|p\|} ( \delta_{ij} p_k + \delta_{jk} p_i + \delta_{ki} p_j  - 2\alpha \|p\| p_k \delta_{ij}
                                           - \alpha \|p\| p_i \delta_{jk} \\ 
                                                    & \hspace{2mm} - \alpha \|p\| p_j \delta_{ik} +  
                                                                \frac{p_i p_j p_k}{\|p\|^2} - 4\alpha \frac{p_i p_j p_k}{\|p\|} )
                                                + (i\hbar )^2 \alpha^2 \left( 2 p_k \delta_{ij} + 3 p_j \delta_{ik} + 3 p_i \delta_{jk} \right)
         \end{array}
\end{equation}
We follow the same level of approximation as in the KMM and DV cases above where terms up to the square of the lowest term in the 
deformation parameter are retained. Next, we have
\begin{equation}
          \begin{array}{ll}
   [x_l, [x_k, [x_i, p_j]]]  =  & (i\hbar)^3 \alpha \left\{ - \frac{1}{\|p\|} (\delta_{ij}\delta_{kl} + \delta_{jk}\delta_{il} + \delta_{ik}\delta_{jl} ) \right.\\
                                             & \hspace{15mm}    +\frac{1}{\|p\|^3} ( \delta_{ij} p_k p_l + \delta_{jk} p_i p_l + \delta_{ik} p_j p_l - \delta_{kl} p_i p_j
                                                                                                - \delta_{jl} p_i p_k - \delta_{il} p_j p_l ) \\
                                             & \hspace{15mm} + \frac{\alpha}{\|p\|^2} (5\delta_{kl} p_i p_j + 5\delta_{jl} p_i p_k + 5\delta_{il} p_j p_k - 
                                                                                         \delta_{ij} p_k p_l  -\delta_{jk} p_i p_l - \delta_{ik} p_j p_l) \\
                                             &  \hspace{15mm}  + \alpha \left. ( 3\delta_{ij}\delta_{kl} + 2 \delta_{jk} \delta_{il} + 2 \delta_{ik}\delta_{jl} ) 
                                                                                                + 3 \frac{p_i p_j p_k p_l}{\|p\|^5} - 11\alpha \frac{p_i p_j p_k p_l}{\|p\|^4} \right\} \\                                                                              
                                             & + (i\hbar)^3 \alpha^2 \left( 2\delta_{ij}\delta_{kl} + 3 \delta_{ik}\delta_{jl} + 3 \delta_{jk}\delta_{il} \right)
          \end{array}
\end{equation}
Calculation of the next few terms such as  \ $[x_m, [x_l, [x_k, [x_i, p_j]]]]$ \ results in a gradually increasing level of complexity of the 
resulting expressions, which does not seem to terminate even in the approximation up to \ $\alpha^2.$ \  The reason behind this behaviour,
which is totally different from that of the KMM and the DV algebras, is not hard to pinpoint: it is the existence of \ $\|p\|$ \ rather than of 
\ $\|p\|^2$ \ and its appearance not only in the numerator, but also in the denominator of (33). As long as it is unclear at this point what is the 
physical principle, if any, dictating the form of (33), and since (33) is not the only expression resulting in an uncertainty relation with desirable 
phenomenological consequences, it may be prudent to avoid the use of \ $\|p\|$ \ itself which introduces these problems for the subsequent 
formalism. One could use instead integer powers of \ $\|p\|$ \ in any generalised algebra, starting its square as in the KMM (9) or DV (26) cases. 
Then this algebra will become nilpotent in the lowest non-trivial approximation in terms of the deformation parameter.\\


\noindent{\large\bf 2.6} \ \  In this subsection, we would like to comment on the terms of (33) involving \ $\|p\|$. \ It seems that the meaning
of this quantity and especially its possible vanishing in the denominator of (33) has not been adequately addressed in the literature of the 
generalised  uncertainty principles, so far. For this reason,  a comment or two may be in order about these issues. The notation and some 
pertinent definitions used in the rest of this Section, can be found in the Appendix and the references cited therein.\\

 We will assume that 
someone works  in the Schwartz space \ $\mathcal{S}(\mathbb{R}^n)$ \ in which the Fourier transform 
\ $\mathcal{F}$ \ is well-defined. Incidentally, it is entirely possible to use another integral transform,  such as the Mellin transform, 
to reach similar conclusions. We immediately see that the symbol corresponding to \ $\| p \|$ \ is 
\begin{equation}
     \tilde{p} (x, y) \  = \ \left( y_1^2 + y_2^2 + \ldots + y_n^2 \right)^\frac{1}{2} 
\end{equation}
This is a classical symbol belonging to the (H\"{o}rmander) class \ $S^1 _{1,0}$. \ In more physical terms, it is a pure 
canonical momentum, being independent of the ``configuration" variables \ $x$. \ The corresponding operator \ $p(x, \partial_x)$ \ is a first 
order pseudo-differential operator belonging to \ $OPS^1 _{1,0}$. \ The Laplacian \ $\nabla^2$ \ on \ $\mathbb{R}^n$ \ is a second order 
elliptic operator,  as it has a positive definite symbol, and we see that (41) can be re-expressed as 
\begin{equation}
p(x, \partial_x) \ = \ \left( \nabla^2 \right)^\frac{1}{2}  
\end{equation}
As such, \ $\| p \|$ \ is well-defined on \ $\mathbb{R}^n$. \ We can re-cast (33) in the slightly different form
\begin{equation} 
   [x_i, p_j] \ = \   i\hbar \left\{ \delta_{ij} - \alpha \| p \| \left( \delta_{ij} + \frac{p_i}{\|p\|} \frac{p_j}{\|p\|}  \right) + 
                                              \alpha^2 \|p\|^2 \left(\delta_{ij} + 3\frac{p_i}{\|p\|}\frac{p_j}{\|p\|} \right)  \right\},  \hspace{2mm} i,j = 1, 2, 3  
\end{equation}
It should be understood that (33) and (43) are not necessarily equivalent as the domains of these operator expressions can be 
different, despite their functional equivalence. The interest in operator domains should not be dismissed out of hand as an exercise 
of purely mathematical interest. Indeed the potential physical importance of operator domains has been investigated for over two decades 
in the context of quantum gravity and non-commutative geometry, in particular as it pertains to issues of topology change ([44], [45] and 
references therein). We will tacitly assume that the discussion takes place in the intersection of the domains of (33) and (43).\\     

Going back to (43) one should notice that a potential problem lies is its ``infra-red" behaviour, namely in the possibility that its denominator 
becomes zero. This is in stark contrast with typical issues in functional spaces, especially where the Fourier transforms involved, that are mostly 
concerned about ``ultra-violet" divergences. A way to deal with this may be to compactify \ $\mathbb{R}^3$ \ to the 3-torus \ $\mathbb{T}^3$ \
and only consider functions that obey periodic boundary conditions as is frequently done in Quantum Physics. Another way is to work with 
principal values of potentially divergent integrals, which is a form of infra-red regularization, as is done in the case of Hilbert transforms and other 
singular integral operators. One can take this path by re-writing the fractional terms of (43) in terms of Riesz transforms as
\begin{equation}           
    [x_i, p_j] \ = \   i\hbar \left\{ \delta_{ij} - \alpha \| p \| \left( \delta_{ij} - \mathcal{R}_i \mathcal{R}_j  \right) + 
                                              \alpha^2 \|p\|^2 \left(\delta_{ij} - 3 \mathcal{R}_i \mathcal{R}_j \right)  \right\},  \hspace{2mm} i,j = 1, 2, 3  
\end{equation}
It is well-known [39]-[41] that the Riesz transforms are bounded in \ $L^p(\mathbb{R}^n), \  1 < p < \infty $. \ Among them, the \ $L^2$-integrable functions 
are of greatest interest in Physics, hence (43) therefore (33) are well-defined in such spaces, which is sufficient for our purposes. \\   


\noindent{\large\bf 2.7} \  \ One issue arising from the ADV algebra (33), (34), due to the presence of the inverse of \ $\|p \|$ \ and its square, 
is related to smoothness. As can be seen from (12) and (42), in (33) we are dealing with an inverse power of the Laplacian, of the specific form
\begin{equation}
\mathcal{I}_s = (\nabla^2)^{-\frac{s}{2}}
\end{equation}
For our purposes \ $s=1$ \ but it does not hurt to be somewhat more general and allow \ $s\in\mathbb{C}$, \ with 
\ $Re \ s >0$ \  for convergence purposes. Such expressions are called Riesz potentials of order \ $s$ \ and are defined via the Fourier transform 
and its inverse, for \ $f\in\mathcal{S}(\mathbb{R}^n)$ \ by    
\begin{equation}
     (\mathcal{I}_s f)(x) \ = \ \frac{1}{2^s \pi^\frac{n}{2}} \frac{\Gamma \left( \frac{n-s}{2} \right) }{\Gamma \left( \frac{s}{2} \right) }
                                                          \int_{\mathbb{R}^n} \ \frac{f(x-y)}{|y|^{n-s}} \ d^ny 
\end{equation}
The effect of operators such as \ $\mathcal{J}_s$ \ is to improve the integrability of functions, namely to map \ $\mathcal{I}_s: \ L^p \rightarrow L^q$ \
which for \  $s\in\mathbb{R}$ \ are related by the Sobolev duality  
\begin{equation}
    \frac{1}{p} - \frac{1}{q} = \frac{s}{n}
\end{equation}
In the particular case of the ADV algebra above \ $p=2, \ s=1, \  n=3$ \  which gives \ $q=6$. \ The Hardy-Littlewood-Sobolev fractional integration 
theorem [39]-[41] provides bounds for the corresponding norms. The Riesz potentials such as \ $\|p\|^{-1}$ \ are essentially integrals, hence they act as 
smoothing operators. Therefore, by using inverse powers of differential operators, the corresponding expressions become more regular, a clearly 
desirable property especially for any theory having a classical limit.\\

 On the other hand, because of the fact that integral operators are defined in domains of \ $\mathbb{R}^n$ \ rather than points, an obvious question arises 
 about the meaning of locality in models using  the ADV algebra. 
This, does not only  create the usual problems of interpretation as in ordinary Quantum Physics but also introduces considerably greater difficulties in the
quantization of any such systems.
It is not clear, to us at least,  what exactly would be the meaning of an operator defined at a point in space, even in the distributional sense, 
when a non-local operation such as the convolution with a singular integral operator is involved in even defining the algebra expressing the dynamics. 
A similar issue is also raised and partially addressed in [6] for the KMM algebra by utilising a generalised Bargmann-Fock representation and by defining 
an approximate ``quasi-position" representation. One could possibly address such issues by following a largely algebraic path, in the context of
generalizations of \ $C^{\star}$ \ algebras emulating the path followed in axiomatic quantum field theory [46] or non-commutative geometry [22],[23].
The issue of locality is of central importance in theories of quantum gravity, such as Loop Quantum Gravity [47] or Causal Sets [48], which aim to formulate  
theories that are background independent. Contrast this with the approach taken toward gravity quantization and interaction unification by the 
String/Brane/M theories [49],[50]. Since a generalised uncertainty principle should reflect, in some part, elements of these quantum 
gravitational theories, its treatment of locality is crucial, extending far beyond the mere technical level that we have alluded to here.      \\  


\noindent{\large\bf 2.8} \ \  In [7],[8]  an inner product was introduced in an attempt to develop a rudimentary representation aspects of the KMM algebra. 
It was given, in $n$-dimensional momentum space, by 
\begin{equation}
     (\psi, \phi) = \int_{\mathbb{R}^n}  \frac{1}{1+\beta p^2}\  \psi^{\ast}(p) \phi(p) d^np
\end{equation}
This product can be seen as ``natural" from the viewpoint of the Fourier transform of the inner product of the (Sobolev) Bessel potential space \ 
$L^2_{-1}(\mathbb{R}^n)$. \  More generally, one can see that the algebras giving rise to the  generalised uncertainty principles, can be 
approximately expressed as ``quantizations" in Hilbert spaces endowed with generalized inner products. Modifying the inner product to bypass 
altogether the Stone - von Neumann theorem, which however does not hold in quantum field theory, and  obtain distinct 
predictions from the usual operator quantisation of the Fourier modes of the phase space variables is one of the tenets of theories 
such as Loop Quantum Gravity [47]. Motivated by the the generalised uncertainty principles  as well as by the approach implemented in  
Loop Quantum Gravity, it may be of some interest  to check on whether the Weyl correspondence [39]
can be modified/extended to apply to the above or any new algebras giving rise to the generalized uncertainty principles.\\           

                                                                                                    \vspace{10mm}


                                                         \centerline{\large\sc 3. \  The ``symplectic camel" and  generalized measure theories.}

                                                                                                    \vspace{5mm}

We saw in the previous section that some of the proposed generalisations of the Heisenberg uncertainty principle lead to solvable, and in 
particular limits, to nilpotent Lie algebras. It may be worth wondering on whether this just a coincidence, whether it is part of a mathematical 
pattern, or even more importantly whether it is a manifestation of a physical principle. This section is an attempt to relate such questions to 
known facts about Classical Mechanics and a generalized measure hierarchy of potential use for Quantum Gravity, as first rudimentary 
comments that may contribute toward an answer.\\

\noindent {\large\bf 3.1} \  \ The operator formalism of Quantum Mechanics has undeniable similarities with the Hamiltonian formulation of Classical 
Mechanics. Then, it would be highly suggestive if an extension of Heisenberg's uncertainty principle could be traced back to the structure of  
Classical Mechanics. The fact that this is indeed possible for the Heisenberg uncertainty principle itself, is a relatively recently established 
fundamental result in Symplectic Topology called the ``symplectic non-squeezing theorem" [24], [25]-[35]  or the principle of the ``symplectic camel" [28]. 
As this result has not yet received the visibility in Physics that it duly deserves, despite the extensive  efforts of primarily M. de Gosson (and collaborators), 
who  seems to be its biggest advocate in the Physics community [29]-[34], we will say a few words about it that are related to the present work.\\    

The following applies to any symplectic manifold \ $\mathcal{M}$ \ but we may wish to think more physically as \ $\mathcal{M}$ \ being the 
phase space of a Hamiltonian system. Assume that \ $dim \mathcal{M} = 2n$ \ and let it be parametrized locally by \ $(x, p)$ \ where \ $x = (x_1, \ldots, x_n)$ \ 
and \ $p = (p_1, \ldots, p_n)$ \ where the notation is borrowed from the Hamiltonian formulation of Mechanics. Consider the ball 
\begin{equation}
B_{2n}(R) \ = \ \left\{ (x,p) \in \mathcal{M}: |x|^2 + |p|^2 \leq R \right\} 
\end{equation}
and the ``cylinder"  \ $Z_l(r), \ l = 1,\ldots, n$ \ over the symplectic 2-plane \ $(x_l, p_l)$ \ given by
\begin{equation} 
     Z_l(r) \ = \ \left\{ (x,p) \in \mathcal{M} : x_l^2 + p_l^2 \leq r \right\} 
\end{equation}   
Consider a (smooth) canonical transformation (symplectomorphism) \ $f: \mathcal{M} \rightarrow \mathcal{M}$. \ 
The symplectic non-squeezing theorem states that it is impossible to fit \ $B_{2n} (R)$ \ inside \ $Z_l(r)$ \ unless \ $R\leq r$, \ namely that   
\begin{equation}
     f(B_{2n} (R)) \subset Z_l(r) \ \iff \ R\leq r
\end{equation}
This shows that a phase-space volume is not only preserved by a Hamiltonian (more accurately: a divergence-free) flow, as given by Liouville's 
theorem, but it possess an additional rigidity associated with its projections along each 2-plane of canonical coordinates. Alternatively, the set of 
canonical transformations of \ $\mathcal{M}$ \  is quite different from the set of volume-preserving transformations of \ $\mathcal{M}$ [24],[26],[28]. \ 
This can be interpreted as a rigidity property of Hamiltonian Mechanics whose Quantum Physics ``analogue" is the Schr\"{o}dinger-Robertson inequality [31]-[34] 
\begin{equation}
       (\Delta x_l)^2 (\Delta p_l)^2 \ \geq \ (Cov(x_l, p_l))^2 + \frac{\hbar^2}{4},  \hspace{10mm} l = 1,\ldots, n
\end{equation}
where \ $Cov(x_l, p_l)$ \ stands for an element of the covariance matrix. If the covariance matrix is zero, this results in the usual Heisenberg uncertainty 
relation. So we see that inside Classical Mechanics itself, there are ``elements" of Quantum Physics, when some terms are properly interpreted. 
Is it possible to use further rigidity results of Classical Mechanics (if any further rigidity exists at all) to guide us in formulating a generalised uncertainty 
principle,  therefore going beyond Quantum Mechanics? This is unclear at present. 
Although a definitive answer is unknown, it appears that there may be additional rigidity properties in the behaviour of 
canonical transformations, appearing in the middle dimension \ $n$ \ as the work of [35] seems to indicate. If such indications are affirmative and more 
rigidity constraints exist for phase-space volumes, nilpotence in this context would be the termination after a finite number of steps of a 
sequence of properly defined involutions of such  rigidity constraints.  \\


\noindent{\large\bf 3.2} \ \ The present work is concerned with nilpotent/solvable associative algebras endowed with a bracket operation, that are 
non-linear generalisations of Lie algebras, and properties of related functional spaces which are the carrier 
spaces of their representations.  It may be worthwhile to see how these ideas may carry over from the canonical to the covariant framework. 
Each of these two approaches 
has its own advantages and limitations, but both provide valuable techniques and insights on how to understand and work out the process of quantisation
in particular models. As is clear from the generalised uncertainty 
principles and the corresponding algebras discussed above, our interest is in uncovering properties related to Quantum Gravity.\\
 
The most striking observation is that it is not really surprising how different 
 is Quantum Mechanics from Classical Mechanics, but how actually close they are to each other [36],[37].  
 An indication for such a close relation was provided by the symplectic ``non-squeezing" theorem discussed above. 
 Another is found if one thinks  about a triple-slit experiment extending Young's double-slit experiment [36],[37]. 
 We start with all three slits open and then gradually start blocking off one, then two at a time and then all three.
 We record the corresponding interference patterns with an overall plus sign if three and one slits are open and with an overall negative sign if two or no slits 
 are open. We superimpose these eight resulting patterns by adding them up algebraically. The result will always be zero. If a four, five etc slit extension of 
 Young's experiment is set up and calculations are performed along similar lines, the result will always turn out to be zero. This is a direct consequence of 
 the fact that the Heisenberg algebra is 2-step nilpotent. In Classical Mechanics no new information beyond the one provided by a ``single-slit" 
 experiment is obtained. In Quantum Mechanics, Young's double slit experiment contains all the non-trivial physical information and every multi-slit 
 experiment beyond it gives nothing new. It is in this sense that Quantum Mechanics is as ``close" to Classical Mechanics as ``possible" [36],[37]
 although, of course, their structures are quite different from each other.\\
 
 To generalize this nilpotentcy in the covariant framework, we have to think in terms of generalised measures of histories, expressing the evolution of a 
 system. Consider a set of histories \ $S_1$ \ having a generalised measure indicated by \ $|S_1|$. \ Consider a second set \ $S_2$ \ and form the 
 disjoint union \ $S_1 \amalg S_2$. \ These two sets could be chosen to represent the histories of the electron going through slit one or only through 
 slit two in Young's double slit experiment. The extension of the notation and the definitions to a multi-slit experiment involving the ``histories"
  \ $S_l, \ l=1,\ldots, n$ \  is immediate. Consider a hierarchy of sum rules [36]-[38] 
 
 \begin{equation}
      \begin{array}{rcl}
     I_1 (S_1) &  \equiv  & |S_1| \\
                      &               &   \\
     I_2 (S_1, S_2) &  \equiv  & |S_1 \amalg S_2| - |S_1| - |S_2| \\
                                &               &        \\
     I_3 (S_1, S_2, S_3) &  \equiv  & |S_1 \amalg S_2 \amalg S_3| - |S_1 \amalg S_2| - |S_2 \amalg S_3| - |S_1 \amalg S_3| + |S_1| + |S_2| + |S_3| \\
                                         &    \vdots  &   \\                                      
     I_n (S_1, S_2, \ldots, S_n) &  \equiv  &  \Big|  S_1 \amalg  \ldots \amalg S_n \Big| \ - \
                                                                               \sum\limits_{l=1}^n \Big| S_1 \amalg \ldots \amalg \not\!{S_l} \amalg \ldots \amalg S_n \Big|  \ + \    \\
                                                     &              &               \sum\limits_{\stackrel{l_1, l_2=1}{l_1 \neq l_2}}^n \Big| S_1 \amalg \ldots \amalg \not\!{S_{l_1}}
                                                                                                          \amalg \ldots\amalg \not\!{S_{l_2}}\amalg\ldots\amalg S_n \Big| + \ldots + 
                                                                                                                         (-1)^{n+1} \sum\limits_{l=1}^n \big| S_l \big|  \\    
      \end{array}
 \end{equation}
 Here \ $\not\!\!{S}$ \ indicates that the argument \ $S$ \ should be omitted in the calculation.
 Evidently \ $I_1 \neq 0$ \ for any non-trivial statement to be feasible. Classical Mechanics corresponds to \ $I_2 = 0$. \ Quantum Mechanics
 is given by \ $I_2 \neq 0, \ I_3 = 0$. \  One can straightforwardly see that \ $I_{l+1} = 0$ \ implies that \ $I_l$ \ is additive in each of its arguments. 
 This multi-additivity can be used to explain why imposing \ $I_3 = 0$ \ results in being able to express the real part of the decoherence functional as 
 \ $I_2(S_l, S_l) = 2 |S_l |$, \ which in turn implies that the transition probabilities are proportional to the square of amplitudes, as is well-known in 
 Quantum Physics. A generalised uncertainty principle would reflect in this framework that \ $I_l \neq 0, \ \ l\geq 4$. \ The generalisation of the 
 Heisenberg algebra to an $l$-step nilpotent algebra would be expressed by demanding that \ $I_{l+1} = 0, \ \ l \geq 4$. \\
 
  In such theories, the transition
 probabilities would be functions of some integral power, but not the square, of the amplitudes of wave-functions. It appears that following the 
 covariant approach would also imply that the carrier spaces of the representations of the generalised algebras would be \ $L^p (\mathbb{R}^n), \ p\neq 2$, \
 if not more general Sobolev spaces. Such spaces of functions are in general Banach spaces, rather than Hilbert spaces like \ $L^2 (\mathbb{R}^n)$ \ 
  which is the one 
  used Quantum Physics. This poses an obvious problem, as the Banach spaces \ $L^p(\mathbb{R}^n), \ p\neq 2$ \ do not admit 
  an inner product. Then one  would have to explain how exactly the geometric structure of the Euclidean spaces stems from that of functions which are 
  elements of \ $L^p(\mathbb{R}^n), \ p \neq 2$. \ This might be feasible by technically utilising a Littlewood-Paley type of treatment [39]-[41], 
  but the physical principle that may justify such a ``semi/classical" transition \ $L^p(\mathbb{R}^n), \ p\neq 2$ \ to \ $L^2(\mathbb{R}^n)$, \ is not clear to us.\\

                                                                                                  \vspace{8mm}
 

                                                                                                   \newpage

                                                             \centerline{\large\sc  4. \ \ Outlook and speculations}              

                                                                                                \vspace{5mm}
    
In this work we attempted to check to what extent some of the, largely, phenomenologically-motivated  generalised uncertainty relations
stem from algebras that are solvable, or nilpotent at least in some approximation. We found that if such proposed algebras
do not contain a simple part that remains unaffected by the Inon\"{u}-Wigner type contraction of one of their deformation parameter(s), 
then they can be seen as being parts of a solvable algebraic structure. 
In appropriate limits of parameters depending on the Planck length and mass, such algebras can be seen to 
possess a nilpotent structure.\\

 It may be worth noticing that the solvable algebras/groups are in a sense complementary to the simple ones that 
we use extensively in various parts in Classical and Quantum Physics. This complemetarity can be seen in two ways: the Killing-Cartan form on solvable
Lie algebras is trivial but it is non-zero for simple algebras. Alternatively, any Lie algebra can be expressed as a semi-direct product 
of a solvable and a semi-simple Lie algebras, according to the Levi-Mal'tsev decomposition.   
We are cannot help but wonder on whether this complementarity  persists at a more fundamental level and has any 
significance for Quantum Gravity  or it is just a formal coincidence due to our treatment and approximations? \\
    
If such a solvability and nilpotency are accepted, then it may be worth examining the form of the generalised measure theories that may be appropriate 
for formulating the corresponding covariant formalism. In our opinion, this raises obvious questions about the central role that the Hilbert spaces play 
in Classical and Quantum Physics. We believe that it may be worth further exploring the physical and formal reasons as well as the corresponding 
implications that may be behind such a role.\\

The ADV algebra also raises some questions that may be of interest: Should we even allow for pseudo-differential and smoothing operators 
in fundamental algebras? If so, what may be implications on locality or on the Markovian character of the classical and quantum evolution?
What techniques could someone use to explore further such ideas? We believe that some of these questions may merit some attention in  
future work.\\  

Lastly, one cannot fail to see the resemblance of (53) to a simplicial structure. It may 
be of interest to explore consequences of such a simplical view, define appropriate 
boundary/co-boundary operators and a (co-)homology theory [51], generalise valuation theory [52] etc. \\  
         
                                                                                    \vspace{2mm}
 
 
                                                      \centerline{\large\sc Acknowledgement} 

                                                                                     \vspace{5mm}

 We are grateful to Professor M. de Gosson for bringing his work to our attention and for sending us a copy of [32]  prior to its publication.\\

                                                                           \vspace{8mm}
                                                                           
 
                                                          \centerline{\large\sc Appendix}
 
                                                                                    \vspace{5mm}
 
 Here, we collect some very well-known facts from harmonic analysis and pseudo-differential operators that may be of some use in reading \
 Subsections 2.6, 2.7. We follow [39]-[42].\\
 
\noindent{\bf I.} \  The Schwartz space \ $\mathcal{S}(\mathbb{R}^n)$ \ is the subspace of smooth functions of \ $C^\infty(\mathbb{R}^n)$ \ such that 
themselves as well as 
 their derivatives decay faster than the inverse of any polynomial at infinity. To be more precise, define the multi-indices \ 
 $\alpha = (\alpha_1, \ldots, \alpha_n), \  \beta = (\beta_1, \ldots, \beta_n)$ \ with \ $\alpha, \ \beta\in \mathbb{N}^n$ \ by 
 \begin{equation}
     x^\alpha = x_1^{\alpha_1} x_2^{\alpha_2} \cdots x_n^{\alpha_n}, \hspace{10mm} 
                                  \partial_x^\beta = \frac{\partial^{\beta_1}}{\partial x_1^{\beta_1}} \frac{\partial^{\beta_2}}{\partial x_2^{\beta_2}} \cdots
                                                                      \frac{\partial^{\beta_n}}{\partial x_n^{\beta_n}}                                  
\end{equation}
with
\begin{equation}
        |\alpha | \ = \  |\alpha_1| + \alpha_2| + \cdots + |\alpha_n|,  \hspace{10mm} |\beta| \ = \ |\beta_1| + |\beta_2| + \cdots + |\beta_n| 
\end{equation}
Consider \ $f: \mathbb{R}^n \rightarrow \mathbb{C}$ \ such that \ $f \in \ C^\infty(\mathbb{R}^n)$ \ and for each pair of multi-indices \ $\alpha, \ \beta$ \ 
define the semi-norms
 \begin{equation}
     \| f \|_{\alpha, \beta} = \sup_{x\in\mathbb{R}^n} |x^\alpha \partial_x^\beta f(x)| 
 \end{equation}
 The above set of denumerable semi-norms allow one to define the Schwartz space by 
 \begin{equation} 
    \mathcal{S} (\mathbb{R}^n) = \{ f\in C^\infty (\mathbb{R}^n): \  \| f \|_{\alpha, \beta} < \infty, \ \ \ \forall \ \alpha, \beta \in \mathbb{N}^n \}
 \end{equation} 
 One can immediately see that an equivalent definition of \ $\mathcal{S}(\mathbb{R}^n)$, \ with \ $C_{\beta, N} > 0$ \ constants is
  \begin{equation}
         |\partial^\beta f (x)| \  \leq \  C_{\beta, N} (1+|x|)^N, \hspace{5mm}  \forall \  \beta\in\mathbb{N}^n, \  \forall \ N  \in \mathbb{N} 
  \end{equation}
 The dual to \ $\mathcal{S}(\mathbb{R}^n)$, \ namely the space of linear functionals on  \ $\mathcal{S}(\mathbb{R}^n)$ \
is indicated by \ $\mathcal{S}^\prime (\mathbb{R}^n)$ \ and is called space of tempered distributions of \ $\mathbb{R}^n$. \
 The Fourier transform for \ $f\in \mathcal{S}(\mathbb{R}^n)$ \ is defined by 
 \begin{equation}
          \mathcal{F}[f] (\xi ) \equiv \hat{f} (\xi) \ = \ \frac{1}{(2\pi)^{\frac{n}{2}}} \int_{\mathbb{R}^n}  f(x) \  e^{-ix\cdot \xi} \ d^n x
 \end{equation}
 and the inverse Fourier transform is 
\begin{equation} 
    \mathcal{F}^{-1}[f] (x) \equiv \check{f} (x) \ = \ \frac{1}{(2\pi)^{\frac{n}{2}}} \int_{\mathbb{R}^n}  \hat{f}(\xi ) \ e^{ix\cdot \xi} \ d^n \xi
\end{equation} 
In the above equations \ $x\cdot \xi$ \ indicates the Euclidean inner product and \ $|x|$ \ stands for  the Euclidean norm of \ $x\in\mathbb{R}^n$. \ 
Both the Fourier and the inverse Fourier transforms are unitary operations (isometries), since according to Parseval's identity
\begin{equation}
    \int_{\mathbb{R}^n} f_1(x) f_2^{\ast} (x) \ d^n x \ = \ \int_{\mathbb{R}^n} \hat{f_1}(\xi) \hat{f_2}^\ast \!\! (\xi) \ d^n \xi    
\end{equation}
where \ $^\ast$ \ indicates the complex conjugation, and it immediately implies Plancherel's formula
\begin{equation}
      |f|^2 \ = \  |\hat{f}|^2 \ = \ |\check{f}|^2
\end{equation}


\noindent{\bf II.} \  Consider the function \ $\tilde{\sigma} (x,y): \mathbb{R}^n \times \mathbb{R}^n \rightarrow \mathbb{C}$. \
For our purposes, it is sufficient to assume that  \ $\tilde{\sigma} \in C^\infty (\mathbb{R}^n \times \mathbb{R}^n)$. \  Consider \ 
$m\in\mathbb{R}, \  \ 0 < \rho, \delta \leq 1$. \ Then \ $\tilde{\sigma}(x,y)$ \ is called a symbol in the (H\"{o}rmander) class 
\ $S^m  _{\rho, \delta}$, \  if for all multi-indices \ 
$\alpha, \beta \in \mathbb{N}^n$ \ there are constants \ $c_{\alpha, \beta}$ \ such that 
\begin{equation} 
        |\partial_x^\beta \partial_y^\alpha \tilde{\sigma} (x,y)| \  \leq \  c_{\alpha, \beta} \ \langle y\rangle ^{m-\rho |\alpha| + \delta |\beta |}
\end{equation}
where 
\begin{equation}
      \langle x \rangle \ \equiv  \  (1+ |x|^2)^\frac{1}{2}
\end{equation}
Consider now the operator \ $\sigma (x, \partial_x): \mathcal{S}(\mathbb{R}^n) \rightarrow \mathcal{S}(\mathbb{R}^n)$ \ given by 
\begin{equation}
      \sigma (x, \partial_x) f(x) \ = \frac{1}{(2\pi )^\frac{n}{2}}  \int_{\mathbb{R}^n} \tilde{\sigma}(x,\xi ) \hat{f}(\xi ) e^{ix\cdot \xi} \ d^n\xi  
\end{equation}
If \ $\tilde{\sigma} (x, y) \in S^m _{\rho, \delta}$, \ then \  $ \sigma (x, \partial_x)$ \ is a pseudo-differential operator belonging to the class 
\ $OPS^m _{\rho, \delta}$. \  In the above definitions, \ $m$ \ is called the order of the operator. If \ $\tilde{\sigma}$ \ is  polynomial, then
the corresponding operator $\sigma$ is differential. If the symbols \ $\tilde{\sigma}(x,y)$ \ 
can be decomposed asymptotically, as sums of homogeneous functions of degrees \ $m-j$, \ namely if
\begin{equation}
    \tilde{\sigma}(x,y) - \sum_{j=0}^N \tilde{\sigma}_{m-j}(x,y) \  \in \ S^{m-N}_{1,0}
\end{equation}
where
\begin{equation}
     \tilde{\sigma} (x,ty) \ = \ t^j \ \tilde{\sigma}(x,y), \hspace{10mm} t\in\mathbb{R}, \  |y| \ \geq \ 1
\end{equation}
then they are called classical symbols. The highest order term in the above classical symbol expansion is called the principal symbol.
An element \ $\sigma\in OPS^m_{\rho, \delta}$ \ is called elliptic pseudo-differential operator, if for some \ $R < \infty$ \ there is a constant 
\ $c>0$ \ such that  
 \begin{equation}  
      |\tilde{\sigma} (x, y) | \ \geq c \  \langle y \rangle ^m, \hspace{8mm} |y| \geq R 
 \end{equation}
 

\noindent{\bf III.} \  Sobolev spaces are spaces of functions aiming to quantify the ``degree of the functions' 
smoothness". First, and as a reminder, one defines the Lebesgue spaces 
\begin{equation}
     L^p (\mathbb{R}^n) \ = \ \{ f: \mathbb{R}^n \rightarrow \mathbb{C}:  \ \int_{\mathbb{R}^n} |f(x)|^p \  d^n x \ < \ \infty \}
\end{equation}
It turns out that these are  Banach spaces when equipped with the \ $L^p$ \ norm
\begin{equation}
             \|f\|_{L^p} \ = \  \left( \int_{\mathbb{R}^n} |f(x)|^p  \ d^n x \right)^\frac{1}{p}  
\end{equation}
For the triangle inequality to hold \ $1\leq p \leq \infty$ \ where \  $L^\infty$ \ is equipped with the sup norm.
The classical Sobolev spaces \  $W^{k,p}(\mathbb{R}^n), \ k, p \in \mathbb{N}$ \ are defined as  
\begin{equation}
    W^{k,p} (\mathbb{R}^n) \ = \ \left\{ f \ \in L^p (\mathbb{R}^n): \ \|f\|_{W^{k,p}}  = \sum_{|\beta | \leq k} \|\partial^\beta f \|_{L^p} < \infty \right\} 
\end{equation}
An alternative description of \ $W^{k,p}(\mathbb{R}^n)$, \ which also allows for an extension to \ $k\in\mathbb{R}$, \ is given via the 
Fourier transform and the Bessel potential spaces
\begin{equation}   
           L^p_k (\mathbb{R}^n ) \ = \ \left\{ f \in L^p (\mathbb{R}^n):  \bigg\| \mathcal{F}^{-1} \left[ (1+ | \xi |^2)^\frac{k}{2}  \mathcal{F}[f] (\xi) \right](x) \bigg\|_{L^p} 
                               < \infty  \right\}
\end{equation}
A theorem of Calder\'{o}n states that for \ $k\in\mathbb{N}$, \ indeed \ $W^{k,p}(\mathbb{R}^n) = L^p_k (\mathbb{R}^n)$. \ Among the above functional 
spaces, the  most commonly used in Physics have, undoubtedly, been \ $L^2(\mathbb{R}^n)$ \ and \ $W^{k,2}(\mathbb{R}^n)$ \ both of which are Hilbert 
spaces. The inner product \ $(\cdot, \cdot)_k$ \ of \ $W^{k,2}(\mathbb{R}^n)$ \ is given in terms of the usual \ $L^2$ \ inner product \ $(\cdot , \cdot)$ \ by 
\begin{equation}
          (f_1, f_2)_k \ = \ \sum_{|\beta| \leq k} \ (\partial^\beta f_1, \partial^\beta f_2)
\end{equation} 
Due to the equivalence of the norms \ $1+|y|$ \ and \ $\langle y \rangle$ \ of \ $L^2$ \ one can extend this to an inner product in 
\ $W^{k,2}, \ k\in\mathbb{R}$ \ by 
\begin{equation}
    (f_1, f_2)_k \ = \  \int_{\mathbb{R}^n} \langle \xi \rangle ^{2k} \ \hat{f_1}(\xi) \ \hat{f_2}\! ^\ast (\xi ) \ d^n \xi  
\end{equation}  
which gives rise to the norm 
\begin{equation}
       \| f \|^2_k \ = \  \int_{\mathbb{R}^n} \langle \xi \rangle^{2k} |\hat{f}(\xi )|^2 \ d^n\xi
\end{equation}
It may be worth observing that if \ $f \in \mathcal{S}(\mathbb{R}^n)$, \ then \ $f\in L^p_k (\mathbb{R}^n), \ k \in\mathbb{R}$. \ 
A pseudo-differential operator, such as \ $\sigma (x, \partial_x) \in OPS^m _{\rho, \delta}$, \ can be extended to an operator acting between 
the Sobolev spaces \ $L^p _{k+m} (\mathbb{R}^n) \rightarrow L^p _k (\mathbb{R}^n)$ \  or on the space of tempered distributions \
$\mathcal{S}^\prime (\mathbb{R}^n)$. \\    


\noindent{\bf IV.} \ Riesz transforms are multi-dimensional analogues of the Hilbert transforms. For \  $\mathbb{R}^n$ \ the Riesz transforms \ 
$\mathcal{R}_l, \ l=1, \ldots, n$ \ are defined to be singular integral operators of convolution type, as follows: Let \ $f \in \mathcal{S}(\mathbb{R}^n)$. \ Then  
\begin{equation}
    (\mathcal{R}_l f) (x) \ = \   \frac{\Gamma (\frac{n+1}{2})}{\pi^\frac{n+1}{2}} \ p.v. \ \int_{\mathbb{R}^n} \frac{x_l - y_l}{|x-y|^{n+1}} \ f(y) \ d^ny 
\end{equation}
where \ $p.v.$ \ indicates the principal value of the integral and \ $\Gamma(x)$ \ is the Euler gamma function.  More explicitly, 
the Riesz transforms can be seen as the convolutions
\begin{equation}  
      (\mathcal{R}_l f) (x) \ = \ (f \star \phi_l) (x) 
\end{equation}
where  \  $\phi_l \in\mathcal{S}^\prime (\mathbb{R}^n),  \ l=1,\ldots,n$ \ are tempered distributions given by the pairing 
\begin{equation}
    \langle \phi_l, h \rangle \ = \ \frac{\Gamma (\frac{n+1}{2})}{\pi^\frac{n+1}{2}} \ \lim_{\epsilon \rightarrow 0} \int_{|x|\geq\epsilon} \
                                                                       \frac{x_l}{|x|^{n+1}} \ h(x) \ d^nx
\end{equation}
for \ $h\in \mathcal{S}(\mathbb{R}^n)$. \ What is of particular interest for our purposes is that the Fourier transform of the Riesz transform is a 
Fourier multiplier, namely that for \ $f\in\mathcal{S}(\mathbb{R}^n)$, \ we have
\begin{equation}  
     (\mathcal{R}_l f)(x) \ = \ \mathcal{F}^{-1} \left[ -i \ \frac{\xi_l}{|\xi |} \ \mathcal{F}[f](\xi)  \right] (x)
\end{equation}

                                                                        \vspace{15mm}
 
                                                                                                                         
                                                        \centerline{\large\sc References}
 
                                                                          \vspace{5mm}

\noindent [1] C.A. Mead, \ \emph{Phys. Rev. D} {\bf 135}, \ 849 \ (1964).\\ 
\noindent [2] D. Amati, M. Ciafaloni, G. Veneziano, \ \emph{Phys. Lett. B} {\bf 216}, \ 41 \ (1989).\\
\noindent [3] K. Konishi, G. Paffuti, P. Provero, \ \emph{Phys. Lett. B} {\bf 234}, \ 276 \ (1990).\\
\noindent [4] M. Maggiore, \ \emph{Phys. Lett. B} {\bf 304}, \ 65 \ (1993).\\
\noindent [5] M. Maggiore, \ \emph{Phys. Lett. B} {\bf 319}, \ 83 \ (1993).\\
\noindent [6] M. Maggiore, \ \emph{Phys. Rev. D} {\bf 49}, \ 5182 \ (1994).\\
\noindent [7] A. Kempf, G. Mangano, R.B. Mann, \ \emph{Phys. Rev. D} {\bf 52}, \ 1108 \ (1995).\\
\noindent [8] A. Kempf, \ \emph{J. Phys. A} {\bf 30}, \ 2093 \ (1997).\\ 
\noindent [9] L.J. Garay, \ \emph{Int. J. Mod. Phys. A} {\bf 10}, \ 145 \ (1995).\\
\noindent [10] M.-T. Jaeckel, S. Reynaud, \ \emph{Phys. Lett. A} {\bf 185}, \ 143 \ (1994).\\ 
\noindent [11] S. Das, E.C. Vagenas, \ \emph{Phys. Rev. Lett.} {\bf 101}, \ 221301 \ (2008).\\
\noindent [12] S. Das, E.C. Vagenas, \ \emph{Can. J. Phys.} {\bf 87}, \ 233 \ (2009).\\
\noindent [13] C. Bambi, \ \emph{Class. Quant. Grav.} {\bf 25}, \ 105003 \ (2008).\\
\noindent [14] A.F. Ali, S. Das, E.C. Vagenas, \ \emph{Phys. Lett. B} {\bf 678}, \ 497 \ (2009).\\
\noindent [15]   A.F. Ali, S. Das, E.C. Vagenas, \ \emph{Phys. Lett. B} {\bf 678}, \ 497 \ (2009).\\
\noindent [16]   M. Bojowald, A. Kempf, \ \emph{Phys. Rev. D} {\bf 86}, \ 085017 \ (2012).\\
\noindent [17]  W. Chemissany, S. Das, A.F. Ali, E.C. Vagenas, \ \emph{JCAP}, \ 1112:017 \ (2011).\\
\noindent [18]  B. Majumder, \ \emph{Phys. Lett. B} {\bf 709}, \ 133 \ (2012).\\
\noindent [19] Y.-W. Kim, H.W. Lee, Y.S. Myung, \ \emph{Phys. Lett. B} {\bf 673}, \ 293 \ (2009).\\
\noindent [20] M. Kober, \ \emph{Phys. Rev. D} {\bf 82}, \ 085017 \ (2010).\\
\noindent [21] V. Husain, D. Kothawala, S.S. Seahra, \ \emph{Generalized uncertainty principles and quantum
                             \hspace*{6mm}  field theory}, \  {\sf arXiv:1208.5761}\\
\noindent [22] A.H. Chamseddine, A. Connes, \ \emph{Fortsch. Physik} {\bf 58}, \ 553 \ (2010).\\
\noindent [23] A.H. Chamseddine, A. Connes, \ \emph{Space-time from the spectral point of view}, \ {\sf arXiv:1008.0985}\\ 
\noindent [24] M. Gromov, \ \emph{Inv. Math.} {\bf 82}, \ 307 \ (1985).\\ 
\noindent [25] I. Ekeland, H. Hofer, \ \emph{Math. Z.} {\bf 200}, \ 355 \ (1989).\\
\noindent [26] H.Hofer, E. Zehnder, \ \emph{Symplectic Invariants and Hamiltonian Dynamics}, \ Birkh\"{a}user, Basel \\
                            \hspace*{6mm} (1994). \\
\noindent [27] C. Viterbo,  \  \emph{Ast\'{e}risque} {\bf 177-178}, \ 345 \ (1989).\\  
\noindent [28] V.I. Arnol'd, \ \emph{Russian Math. Surveys} {\bf 41:  6}, \ 1 \ (1986).\\                          
\noindent [29] M. de Gosson, \ \emph{J. Phys. A} {\bf 34}, \ 10085 \ (2001).\\ 
\noindent [30] M. de Gosson, \ \emph{J. Phys. A} {\bf 35}, \ 6825 \ (2002).\\
\noindent [31] M. de Gosson, F. Luef, \ \emph{Phys. Rep.} {\bf 484}, \ 131 \ (2009).\\
\noindent [32] M. de Gosson, \ \emph{Found. Phys.} {\bf  39}, \ 194 \ (2009).\\
\noindent [33] M. de Gosson, \ \emph{The Symplectic Camel and Quantum  Universal Invariants: the Angel of \\
                              \hspace*{6mm} Geometry vs. the Demon of Algebra}, \ {\sf arXiv: 1203.5310}\\
\noindent [34] M. de Gosson, \ \emph{The symplectic egg}, \ {\sf arXiv:1208.5969}\\
\noindent [35] A. Abbondandolo, S. Matveyev, \ \emph{How large is the shadow of a symplectic ball?}, \ {\sf  arXiv:1202.3614}\\
\noindent [36] R.D. Sorkin, \ \emph{Mod. Phys. Lett. A} {\bf 9}, \ 3119 \ (1994).\\
\noindent [37] R.D. Sorkin, \ \emph{Quantum Measure Theory and Its Interpretation}, \ in \ \emph{Proceedings of the \\ 
                              \hspace*{6mm} Fourth Drexel Symposium on Quantum Non-integrability: Quantum Classical \\    
                              \hspace*{6mm} Correspondence}, D.H. Feng, B.-L. Hu (Eds.), International Press, Cambridge, Mass. (1997)\\ 
\noindent [38] R. B. Salgado, \ \emph{Mod. Phys. Lett. A} {\bf 17}, \ 711 \ (2002).\\
\noindent [39] E.M. Stein, \ \emph{Harmonic Analysis: Real Variable Methods, Orthogonality and Oscillatory \\
                              \hspace*{6mm} Integrals}, \ Princeton University Press, \ Princeton \  (1993).\\ 
\noindent [40] L. Grafakos, \ \emph{Classical Fourier Analysis}, \ 2nd Ed., \ Springer \ (2008).\\ 
\noindent [41] L. Grafakos, \ \emph{Modern Fourier Analysis}, \ 2nd Ed., \ Springer \ (2009).\\
\noindent [42] M. Taylor, \ \emph{Pseudodifferential Operators and Nonlinear PDE}, \ Birkh\"{a}user, Boston (1991).\\
\noindent [43] S. Helgason, \  \ \emph{Differential Geometry, Lie Groups, and Symmetric Spaces}, \ \  Amer. \ Math.\\
                           \hspace*{6mm}  Soc., \  Providence \ (2001).\\
\noindent [44]  M. Asorey, A.P. Balachandran, G. Marmo, I.P. Costa e Silva, A.R. de Queiroz, P. Teotonio-
                              \hspace*{6mm} Sobrinho, S. Vaidya,  \emph{Quantum Physics and Fluctuating Topologies: Survey},  {\sf arXiv:1211.6882}\\
\noindent [45] A.D. Shapere, F. Wilczek, Z. Xiong, \  \emph{Models of Topology Change}, \ {\sf arXiv:1208.3841}\\
\noindent [46] R. Haag, \ \emph{Local Quantum Physics}, \ 2nd Ed., \ Springer-Verlag, \ Berlin \ (1992).\\
\noindent [47] C. Rovelli, \ \emph{Class. Quant. Grav.} {\bf 28}, \ 153002 \ (2011).\\
\noindent [48] S. Surya, \ \emph{Directions in Causal Set Quantum Gravity}, \  {\sf arXiv:1103.6272}\\
\noindent [49] S. Mukhi, \ \emph{Class. Quant. Grav.} {\bf 28}, \ 153001 \ (2011).\\ 
\noindent [50] M. Blau, S. Theisen,\ \emph{Gen. Rel. Grav.} {\bf 41}, \ 743 \ (2009).\\ 
\noindent [51] E. Spanier, \ \emph{Algebraic Topology}, \ Springer, \ New York \ (1966).\\
\noindent [52] D.A. Klain, G-C. Rota, \ \emph{Introduction to Geometric Probability}, \ Cambridge University\\
                               \hspace*{6mm}  Press, \ Cambridge \  (1997).    \\

\end{document}